\documentstyle[prl,aps,twocolumn,psfig]{revtex}
\begin{document}
\twocolumn[\hsize\textwidth\columnwidth\hsize\csname
  @twocolumnfalse\endcsname
\vspace{-0.5in}
\draft
\title{Using kaon regeneration to probe the quark mixing 
parameter $\cos{2 \beta}$ in $B \rightarrow \psi K$ decays}
\author{Helen R.\ Quinn,
        Thomas Schietinger,
        Jo\~ao P.\ Silva\cite{ISEL},
        and Arthur E.\ Snyder}
\address{Stanford Linear Accelerator Center, 
         Stanford University, 
         Stanford,
	 CA 94309, 
         USA}
\date{\today}
\maketitle
\begin{abstract}
We suggest a novel method to determine the sign
of $\cos {2 \beta}$ in the decays $B \rightarrow \psi K$,
by creating interference between $K_L$ and $K_S$ final states
via ``regeneration,''
that is propagation through a matter target region to convert some
$K_L $ to $K_S$.
The determination of this quantity resolves an ambiguity between 
$\beta$ and $90^\circ -\beta$ that remains after the standard measurements
of $\sin{2 \beta}$ and may turn out to be important in resolving
whether the result is in agreement with Standard Model predictions 
or indicates the presence of new physics.
We find the measurement is feasible at a $B$-factory, 
but requires several years of high-luminosity running with a regeneration 
target affecting a significant fraction of the detector.
\end{abstract}

] 


The determination of the parameters of the quark-mixing 
or Cabibbo--Kobayashi--Maskawa (CKM) \cite{CKM} matrix
as a consistency test of the Standard Model is a 
priority in current high-energy physics research.
The {\sc BaBar} \cite{BaBar} and Belle \cite{Belle} collaborations 
have announced preliminary results from the study of the CP-violating 
asymmetry in the decay $B^0\rightarrow \psi K_S$ \cite{psi} which attain
a precision comparable to the earlier CDF measurement \cite{CDF}.
Eventually, these and other experiments will provide us with an 
accurate and clean determination of the CKM-parameter $\sin 2\beta$.
Unfortunately, knowledge of $\sin 2\beta$ only determines the
angle $\beta$ up to a four-fold ambiguity, meaning that new physics
could be hiding under an apparent confirmation of the Standard Model
\cite{discrete,BLS}.
A measurement of the sign of $\cos 2\beta$ would remove the
$\beta\rightarrow 90^\circ -\beta$ ambiguity.

The interference between the $K_S$ and the $K_L$ decays to 
$\pi\pi$ in principle provides enough information in the time-dependence
of the decay chain $B\rightarrow \psi K \rightarrow \psi (\pi\pi)_K$
to extract $\cos 2\beta$.
However, the CP suppression of the $K_L \rightarrow \pi \pi$ decay
makes such a measurement unrealistic.
Kayser has suggested the use of semileptonic kaon decays to avoid the
CP suppression, but this again leads to tiny effects because of the
small $K_S \rightarrow \pi \ell \nu$ branching ratio \cite{Kay97}.

Here we propose to utilize neutral kaon regeneration to enhance the
$K_S$--$K_L$ interference, leading to a measurement of $\cos 2\beta$
in the decay $B\rightarrow \psi K \rightarrow \psi (\pi\pi)_K$
that does not suffer from CP suppression nor from small branching
ratios.
Regeneration hinges on the fact that the eigenstates in matter
and in vacuum do not coincide.
They are related through $K^\prime_L \sim K_L + r K_S$
and $K^\prime_S \sim K_S - r K_L$,
where
\begin{equation}
r = -\, \frac{\pi N}{m_K} \, \frac{\Delta f}{\Delta\lambda}
\end{equation}
is the regeneration parameter.
Here, $N$ is the density of scattering centers,
$\Delta f = f - \bar{f}$, 
$f$ ($\bar f$) is the elastic forward scattering amplitude of
$K^0$ ($\overline{K^0}$),
whose imaginary part is related by the optical theorem
to the total cross section $\sigma_T$ ($\bar \sigma_T$), 
$\Delta\lambda = \Delta m_K - i\Delta\Gamma_K/2  = 
\lambda_L - \lambda_S$, and
$\lambda_n = m_n -i\Gamma_n/2$ are the vacuum eigenvalues
corresponding to the two kaon mass eigenstates with masses $m_n$ and
widths $\Gamma_n$ for $n=S$ and $n=L$.
We use the approximation $m_K = (m_L + m_S)/2$.

Consider the situation depicted schematically in Fig.~1,
where an initial $| B^0 \rangle$ state evolves at proper $t_B$ as
\begin{eqnarray}
&&
e^{\Gamma_B t_B /2}\, 
| B^0(t_B) \rangle =
\cos{\left( \Delta m_B\, t_B/2 \right)}\,  | B^0 \rangle
\nonumber\\
&&
+
q_B/p_B\ 
\sin{\left( \Delta m_B\, t_B/2 \right)}\,  | \overline{B^0} \rangle .
\label{time-devel-B}
\end{eqnarray}
Here
$\Delta m_B$ is the difference between the masses of 
the two $B$ mass eigenstates $B_H$ and $B_L$,
$\Gamma_B$ is their average width (the difference can be
neglected), and $q_B/p_B$ is defined via the relation 
$|B_{H,L} \rangle = p_B\, | B^0 \rangle 
                \pm q_B\, | \overline{B^0} \rangle$.
At time $t_B$ the $B$ mixture decays into a $\psi$ and a kaon.
\begin{figure}[tbh]
\centerline{\psfig{figure=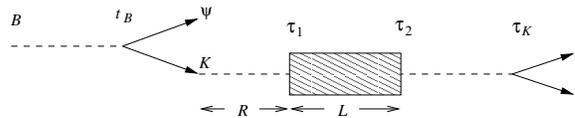,height=0.6in}}
\caption{The diagram illustrates the $B^0 \rightarrow
\psi\, f_K$ evolution when the kaon traverses a
regenerator before decaying into $f_K$.
The time evolutions are indicated by dashed lines and the decays
by solid lines.
\label{fig:1}}
\end{figure}%
The resulting kaon state is given by
$| K_{{\rm from }\ B^0(t_B)} \rangle
= 
\langle \psi K^0 | T | B^0(t_B) \rangle\ |K^0\rangle
+
\langle \psi \overline{K^0} | T | B^0(t_B) \rangle\ 
|\overline{K^0}\rangle$,
which may be rewritten as 
$\alpha_S^0 |K_S\rangle +  \alpha_L^0 |K_L\rangle$,
with \cite{BLS,Kay97}
\begin{eqnarray}
&&
\sqrt{2}\, \alpha_{L,S}^0
=
A(B^0 \rightarrow \psi K^0)\, e^{- \Gamma_B t_B/2}
\times
\nonumber\\
&&
\left[
\cos{\left(\Delta m_B\, t_B/2\right)}
e^{i \beta}
\pm i
\sin{\left(\Delta m_B\, t_B/2\right)}
e^{- i \beta}
\right].
\end{eqnarray}
We have used Eq.~(\ref{time-devel-B}),
written the kaon states in terms of the mass eigenstates,
and used
$- q_B/p_B\ 
A(\overline{B^0} \rightarrow \psi \overline{K^0})/
A(B^0 \rightarrow \psi K^0)\ 
p_K/q_K = \exp(- 2 i \beta)$ \cite{explain}.

We now consider the evolution of this kaon state.
The short- and long-lived components evolve independently until
the kaon hits the regenerator at kaon proper time $\tau_1$,
at which time
$\alpha_n (\tau_1) = \exp{(-i \lambda_n \tau_1)}\, \alpha_n^0$,
with $n=S,L$.
These components mix as the kaon state evolves through matter,
until proper time $\tau_2 = \tau_1+\delta \tau$ when the kaon
emerges from the regenerator of length $L$.
The kaon components at that instant may be written as
\cite{regtheor,Fet96,Thomas},
\begin{equation}
\alpha_i(\tau_2) = e^{- N (\sigma_T + \bar \sigma_T) L/4}
e^{- \Gamma_S\, \delta \tau/2
}
\sum_j m_{ij} \alpha_j(\tau_1),
\end{equation}
where $i,j=S,L$.
To linear order in $r$,
we have $m_{SS} \sim 1$,
$m_{SL} \sim m_{LS} \sim
r\, \left[ \exp{(-i \Delta \lambda\, \delta \tau)} - 1 \right]$,
and $m_{LL} \sim \exp{(-i \Delta \lambda\, \delta \tau)}$.
Thenceforth the two components evolve independently again
according to
$\alpha_n (\tau) = \exp{[-i \lambda_n (\tau - \tau_2)]}\, \alpha_n (\tau_2)$.
At proper time $\tau_K$ the kaon state decays into $f_K$.

Factoring out the $K_S$ lifetime and decay rate,
we obtain that the total decay rate 
(which we denote by $\Gamma_\beta (t_B,\tau_K)$)
is given by the expected normalization factors
$1/2$ $e^{- \Gamma_B t_B}$
$\Gamma[B^0 \rightarrow \psi K^0]$
$e^{- N(\sigma_T + \bar \sigma_T)L/2}$
$e^{- \Gamma_S \tau_K}
\Gamma[K_S \rightarrow f_K]$
multiplied by
\begin{eqnarray}
&&
|a_{SS} + \eta a_{LS}|^2
\left[ 1 - \sin{2 \beta} \sin{(\Delta m_B\, t_B)}\right]
\nonumber\\
&&
+ |a_{SL} + \eta a_{LL}|^2
\left[ 1 + \sin{2 \beta} \sin{(\Delta m_B\, t_B)}\right]
\nonumber\\
&&
+ 2 {\mathrm Im} \left[
(a_{SS} + \eta a_{LS})(a_{SL} + \eta a_{LL})^\ast
\right]
\cos{2 \beta}
\sin{(\Delta m_B t_B)}
\nonumber\\
&&
+ 2 {\mathrm Re} \left[
(a_{SS} + \eta a_{LS})(a_{SL} + \eta a_{LL})^\ast
\right]
\cos{(\Delta m_B t_B)}.
\label{master}
\end{eqnarray}
We have defined
$\eta = A(K_L \rightarrow f_K)/A(K_S \rightarrow f_K)$,\linebreak
$a_{SS}$ = $m_{SS}$,
$a_{SL}$ = $m_{SL} \exp(-i\Delta\lambda\,\tau_1)$,
$a_{LS}$ = $\exp [-i\Delta\lambda\,(\tau_K$$-$$\tau_2)] m_{LS}$,
and 
$a_{LL}$ = $\exp[-i\Delta\lambda  
(\tau_K$$-$$\tau_2$)]\linebreak 
$m_{LS} \exp(-i\Delta\lambda\,\tau_1$).
The second (first) sub-index in $a_{ij}$ indicates
the vacuum eigenstate before the kaon state reaches
(after it leaves) the regenerator.
Eq.~(\ref{master}) and the relations between $a_{ij}$
and $m_{ij}$ are valid for any value of $r$.
However,
when $r$ is large we must use the complete expressions
of the $m_{ij}$ in terms of $r$,
which can be found in Ref.~\cite{Thomas}.
A similar expression applies for $\overline{B}$ decays, 
$\overline{\Gamma}_\beta (t_B,\tau_K)$, with the sign of all 
$\sin(\Delta m_B t_B)$ and $\cos(\Delta m_B t_B)$ terms
reversed.

Kayser \cite{Kay97} has considered the case without a regenerator.
In that case,
$a_{SL}$, $a_{LS}$, $L$, and $\delta \tau$ vanish.
The first line in Eq.~(\ref{master}) is proportional to
$e^{-\Gamma_S \tau_K} \Gamma[K_S \rightarrow f_K]$;
it arises from the decay path 
$B^0 \rightarrow \psi K_S \rightarrow \psi f_K$.
It shows how one measures $\sin{2 \beta}$ in the CP-violating
asymmetry of $B^0 \rightarrow \psi K_S$.
The second line, proportional to
$e^{-\Gamma_L \tau_K} \Gamma[K_L \rightarrow f_K]$,
arises from the decay path 
$B^0 \rightarrow \psi K_L \rightarrow \psi f_K$ and
shows that the CP-violating asymmetry in $B^0 \rightarrow \psi K_L$
has the opposite sign to the asymmetry in $B^0 \rightarrow \psi K_S$.
At intermediate times these two paths interfere and one has access
to the $\cos{2 \beta}$ term on the 
third line,
which in vacuum is given by
$2 |\eta| \exp{(-\Delta \Gamma_K \tau_K/2)} 
\sin{(\Delta m_K \tau_K - \arg{\eta})}$
\cite{Kay97}.
This interference vanishes with $\eta$ and is maximized for semileptonic
decays, where $|\eta| \approx 1$.
A simple estimate shows that measuring $\cos{2 \beta}$ in
$B \rightarrow \psi (\pi \ell \nu)_K$ should require around
a few hundred to a thousand times as many events as
the measurement of $\sin{2 \beta}$ from $B \rightarrow \psi K_S$
\cite{BLS,Kay97}.

One advantage of using a regenerator is that the interference
term exists even when $\eta=0$.
Indeed, in that case the third line
of Eq.~(\ref{master}) becomes proportional to
$2 |r| \left[\exp{(- \Delta \Gamma_K \tau_2/2)} \right. 
\sin{(\Delta m_K \tau_2 - \arg{r})} \linebreak
- \exp{(- \Delta \Gamma_K \tau_1/2)}
\left. \sin{(\Delta m_K \tau_1 - \arg{r})}\right]$,
which determines the asymmetry arising from the $\cos 2 \beta$ term.
Since $\tau_2 - \tau_1 = L m_K/p_K$
(where $p_K$ is the momentum of the kaon),
this term is related to the matter absorption
$\exp{[- N(\sigma_T + \bar{\sigma}_T)L/2]}$ which
determines the rate.
To achieve the maximum sensitivity to $\cos 2 \beta$ we must
find the optimal balance between the two effects.

The observable of interest is the decay rate asymmetry 
\begin{equation}
A_\beta (t_B,\tau_K) = 
\frac{\overline{\Gamma}_\beta (t_B,\tau_K) - \Gamma_\beta (t_B,\tau_K)}
    {\overline{\Gamma}_\beta (t_B,\tau_K) + \Gamma_\beta (t_B,\tau_K)} ,
\end{equation}
shown in Fig.~2 for a kaon momentum of 2 GeV/$c$ and $t_B$ fixed at
the mean $B$ lifetime, in comparison with the asymmetry obtained with
Kayser's semileptonic method,
where the rate is very small.
\begin{figure}[twocolumns]
\centerline{\psfig{figure=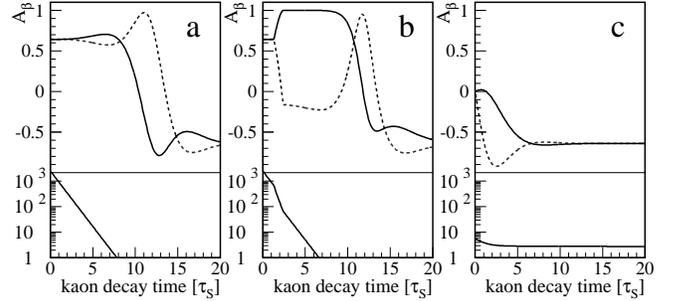,height=1.7in}}
\caption{$B^0 - \overline{B^0}$ decay rate asymmetries in the 
$\psi K$ channel as a function of the kaon decay time
for various kaon decay final states:
(a) $2\pi$;
(b) $2\pi$ with a tungsten regenerator 12~cm thick at a distance of 
14~cm;
and (c) $\pi^-\ell^+\nu$ (Kayser method).
The solid (dashed) lines mark the asymmetries for $\beta = 70^\circ$
($20^\circ$).
A kaon momentum of 2~GeV/$c$ is assumed.
To illustrate the statistical requirements we show in the lower
thirds of the figures the number of expected decays per $\tau_S$ bin 
for 10000 $B\rightarrow \psi K$ decays 
\label{fig:2}}
\end{figure}
It is convenient to integrate the decay rates behind the regenerator as
the dependence on the kaon decay time contains little information on 
$\cos{2 \beta}$ compared to the regeneration effect. 
Neglecting the small contributions from $K_L\rightarrow 2\pi$ decays, 
the asymmetry simply becomes
\begin{equation}
A_\beta^{\text{reg}} (t_B) = 
\frac{\overline{\Gamma}_\beta (t_B,\tau_2) - \Gamma_\beta (t_B,\tau_2)}
    {\overline{\Gamma}_\beta (t_B,\tau_2) + \Gamma_\beta (t_B,\tau_2)} .
\end{equation}
We assume that $\sin{2 \beta}$ will be known with good precision
by the time such an experiment will be considered.
The experimental task will then consist in determining which of the two 
hypotheses, $\beta$ or $90^\circ - \beta$, is favored by the data.
To compare the probabilities of the two hypotheses we calculate the 
logarithm of the likelihood ratio \cite{Orear},
\begin{eqnarray}
{\log \mathcal{R}}(t_B) dt_B &=& 
\frac{1}{2}
n_\beta(t_B) dt_B
\sum_{k=\pm 1}
(1+kA_\beta^{\text{reg}}(t_B))
\nonumber\\
& &  
\times \log \left(\frac{1+kA_\beta^{\text{reg}}(t_B)}
                  {1+kA_{90^\circ-\beta}^{\text{reg}}(t_B)} \right)
\end{eqnarray}
where $n_\beta(t_B) dt_B$ denotes the total number of $K\rightarrow 2\pi$
decays behind the regenerator for which the $K$ originates from a $B$ 
that decayed in the time interval $[t_B,t_B+dt_B]$.
Integration over the $B$ decay time yields the average likelihood ratio:
\begin{equation} \label{average}
\langle {\log \mathcal{R}} \rangle = 
\Gamma_B \int_0^\infty e^{-\Gamma_B t_B} {\log \mathcal{R}} (t_B) dt_B
\end{equation}
The average likelihood ratio is a good measure of the expected 
discrimination power of an experiment and therefore may be used as a 
figure of merit to optimize the regenerator geometry.

To obtain a first estimate of the number of events needed to 
establish the sign of $\cos{2\beta}$ we calculate 
$\langle {\log \mathcal{R}} \rangle$ for a tungsten regenerator 
of varying thickness at varying distance.
We propose tungsten for this experiment because it gives large
regeneration effects.
Since the elastic forward scattering amplitudes $f$ and $\bar{f}$ for 
$K^0$ and $\overline{K^0}$ in tungsten are not known,
we use an empirical scaling law as determined by Gsponer
and collaborators from their measurements in C, Al, Cu, Sn, and Pb for
kaon momenta between 20 to 150~GeV/$c$.
They found that $(\sigma_T+\bar{\sigma}_T)/2 \approx \sigma_T(K_L)$
= 23.5 mb ($A$/g mol$^{-1}$)$^{0.840}$
\cite{Gsp79sum} and $|\Delta f|$ = 1.13 fm ($A$/g mol$^{-1}$)$^{0.758}$ 
($p_K$/GeV$c^{-1}$)$^{0.386}$ \cite{Gsp79diff,Roe77}.
The last result exhibits a power law momentum dependence
in accordance with Regge theory \cite{Regge},
which also predicts that $\arg{\Delta f}$ should be constant and
given by $-(1+0.386)\pi/2 = -0.693 \pi$ \cite{Roe77}.
As a result,
$r$ = 0.033 $e^{i 0.049 \pi}$ ($\rho$/g cm$^{-3}$)
($A$/g mol$^{-1}$)$^{-0.242}$ ($p_K$/GeV$c^{-1}$)$^{0.386}$ 
where $\rho$ is the density of the material used.
The power-law approximation is fairly good down to a few GeV/$c$ momentum,
where low-energy resonances set in \cite{CPLEAR_disp}.
Such resonances can enhance or degrade regeneration effects, but are
not expected to alter significantly the conclusions of this work.
At an asymmetric $e^+e^-$ collider operating at the 
$\Upsilon(4S)$ resonance the average kaon momentum from $B\rightarrow
\psi K$ decays lies around 2~GeV/$c$, so that 
$r$ = 0.24 $e^{i 0.049 \pi}$ for tungsten.

Using these numbers we find that for $\beta = 0^\circ$ ($\sin{2\beta} = 0$), 
$\langle {\log \mathcal{R}}\rangle$ 
has a rather broad maximum for regenerator distances around
$R_{\text{opt}}$ = 17.3 cm and thicknesses of about $L_{\text{opt}}$ = 11.9 cm.
For this case, $\langle {\log \mathcal{R}}\rangle$ equals 0.006\linebreak 
\vspace{-4mm}%
\begin{table}[hbt] 
\caption{Optimal distance and thickness of a tungsten regenerator
         for the geometry depicted in Fig.~1, and for various kaon 
         momenta. Also shown is the average total number of 
         $B\rightarrow \psi K$ events (with the kaon aiming at the 
         regenerator) needed to achieve a separation
         of the $\pm \cos{2\beta}$ hypotheses corresponding to 3$\sigma$.}
\begin{tabular}{rrrrrrr} 
           & \multicolumn{3}{c}{$\beta =  0^\circ$}
           & \multicolumn{3}{c}{$\beta = 70^\circ$} \\
\hline
 $p_K$  & $R_{\text{opt}}$ & $L_{\text{opt}}$ & $N(3\sigma)$ &
          $R_{\text{opt}}$ & $L_{\text{opt}}$ & $N(3\sigma)$  \\
 GeV/$c$   & cm & cm &  & cm & cm &  \\ 
\hline
  1 &  7.1 &  9.4 &  680 &  4.9 &  7.7 & 1230 \\
  2 & 17.3 & 11.9 &  980 & 13.7 & 10.1 & 1670 \\
  5 & 49.5 & 14.4 & 2070 & 44.4 & 12.7 & 3200 \\
 10 &  105 & 15.5 & 4200 &  99  & 14.2 & 6000 \\
 50 &  570 & 16.3 &  26k & 570  & 15.7 & 34k  \\
100 & 1190 & 16.3 &  60k & 1180 & 15.9 & 75k  \\
\label{tab:result}
\end{tabular} 
\end{table}%
\hspace{-3.5mm}%
times the total number of (flavor-tagged) $B\rightarrow \psi K$ decays 
with the $K$ aiming at the regenerator.
This means
that in order to get a likelihood ratio of 370 
(corresponding to a 3$\sigma$ separation) one would need about 980 such 
events.
If $\beta=70^\circ$ ($\sin{2\beta} = 0.7$), the optimum position and
thickness are not much different, but 
$\langle {\log \mathcal{R}}\rangle$ drops
by a factor of 1.7.
Table~1 contains more examples at different kaon momenta
and can be used to estimate the requirements at various
facilities.
At high momenta regeneration effects become smaller and more events
are needed to probe $\cos{2\beta}$. 

A more realistic estimate of the statistical requirements needs to take 
into account experiment-specific details such as the geometry of the detector
and the phase space occupied by the kaons from the $B$ decays.
As an example, we have modeled regenerator geometry as a cylinder
in the {\sc BaBar} detector.

At {\sc BaBar}, 9~GeV electrons collide with 3.1~GeV positrons, copiously 
producing $B\overline{B}$ pairs via the $\Upsilon(4S)$ resonance.
The tight constraints between the kaon momentum and
the polar angle due to the kinematics of the decay chain
$\Upsilon(4S) \rightarrow  B \overline{B}$ followed
by $B \rightarrow \psi K$ 
allow us to optimize the position and thickness of the regenerator.
In this experiment one $B$ decays to a flavor-identifying mode,
or tag, and the other to $\psi K$.
In this case Eq.~(\ref{average}) becomes 
$\langle {\log \mathcal{R}} \rangle = 
\Gamma_B \int_{-\infty}^\infty e^{-\Gamma_B |t|} 
{\log \mathcal{R}} (t) dt$,
where $t = t_{\psi K} - t_{\text{tag}}$ is the time between the two decays.
The value $\langle \log \mathcal{R} \rangle$ is maximal 
for the 9 cm thick cylindrical regenerator 
placed at a radius of 5 cm and a polar angle of $120^\circ$.
At this point each kaon aimed at the regenerator contributes 
0.008 to $\langle \log \mathcal{R} \rangle$ which implies that
about 600 perfectly tagged events would achieve a
3$\sigma$ separation between $\beta=0^\circ$ and $\beta=90^\circ$.
For a regenerator occupying 25\% of the 
instrumented solid angle around the optimal point,
we estimate that (with a merged $J/\psi$ and $\psi(2S)$ sample)
a luminosity of about 300 fb$^{-1}$ would be needed to reach 
3$\sigma$ separation, where we have used the reconstruction 
efficiency and effective tagging efficiency currently achieved by 
{\sc BaBar} \cite{BaBar}
and assumed that tagging particles that hit the regenerator are lost.
If $\beta = 70^\circ$ then about 600 fb$^{-1}$ would be needed.

We have also briefly considered the application of this method at
a hadron collider experiment where hundreds of thousands of reconstructed
$\psi K$ decays and an effective tagging efficiency of $\approx$10\%
are expected.
However, this statistical advantage is offset by the reduction in the 
size of the regeneration effect,
because the difference between the $K^0$ and the $\overline{K^0}$
cross sections drops with momentum (see Table~1)
and because the broad spectrum of kaon momentum makes it impossible
to design a regenerator optimal for all decays.
A regenerator optimized for 2~GeV$/c$
averages about 38\% of its optimum for momenta 
between 1 and 10~GeV/$c$.
So, 
an experiment carried out in the central region where kaon momenta
are modest might be more effective. 

In passing we would like to point out that a tungsten target inside 
the detector would not only act as a regenerator, but also as a strangeness
indicator via reactions such as 
$\overline{K^0} n\rightarrow K^- p$ or
$\overline{K^0} p\rightarrow \Lambda\pi^+$ etc.
The outgoing strange particle unambiguously identifies the strangeness
of the neutral kaon at the time of the interaction in the regenerator,
just as the lepton from a semileptonic decay tags the strangeness at 
the time of decay, allowing us to apply Kayser's method with considerably
higher statistics.
Comparing Figs. 2b and 2c we see that an optimized regenerator would
also be in the right position for this method to be sensitive to $\cos
2\beta$.
Depending on the capability of the detector to reconstruct and identify
such events, combined analysis of the two effects may substantially 
reduce the number of events needed to determine the sign of $\cos 2\beta$.

If the measurement of the sign of $\cos 2\beta$ becomes an issue of
significant interest, then the method suggested here offers an 
opportunity to measure it at a high-luminosity $B$-factory.
It requires insertion of a tungsten target deep inside the detector,
affecting a significant portion of the detector solid angle.
Thus it is not without cost for other physics measurements,
so it will be a question of competing priorities to decide
whether to implement this approach.
Similar considerations will apply in a hadronic $B$-physics experiment,
but detailed simulation of the actual situation is needed before one
can say which configuration of machine,
target and detector offers the best choice for this measurement.

This work was supported by the U.S. Department of Energy 
under contract DE-AC03-76SF00515.
The work of J.\ P.\ S.\ is supported in part by Fulbright,
Instituto Cam\~oes, and by the Portuguese FCT, under grant
PRAXIS XXI/BPD/20129/99	and contract CERN/S/FIS/1214/98.


\begin{references}
\bibitem{psi}
We use $\psi$ to denote both $J/\psi$ and $\psi(2S)$.
\bibitem{explain}
Since the $\bar b \rightarrow \bar c c \bar s$ decays are dominant,
one expects that most models beyond the SM will not affect 
this decay dramatically.
They might, however,
modify the loop-induced $B^0 - \overline{B^0}$ mixing,
through a new mixing phase $\theta_d$.
Under these circumstances,
$\exp(- 2 i \beta)$ should be replaced with
$\exp[- 2 i (\beta + \theta_d)]$.
\end{references}
\end{document}